\documentclass[useAMS,usenatbib,usegraphicx]{mn2e}
\def\LB1{L1157--B1}

\usepackage{fixltx2e}
\title[Acetaldehyde in L1157--B1]{Astrochemistry at work in the L1157--B1 shock: acetaldehyde formation}  
\author[F. Codella et al.]{C. Codella$^{1}$\thanks{E-mail:
codella@arcetri.astro.it }, F. Fontani$^{1}$, 
C. Ceccarelli$^{2,3}$, L. Podio$^{1}$, S. Viti$^{4}$,  
R. Bachiller$^{5}$, \newauthor
M. Benedettini$^{6}$, B. Lefloch$^{2,3}$
\\
\\
$^{1}$ INAF-Osservatorio Astrofisico di Arcetri, L.go E. Fermi 5, Firenze, 50125, Italy \\
$^{2}$ Univ. Grenoble Alpes, IPAG, F-38000 Grenoble, France \\
$^{3}$ CNRS, IPAG, F-38000 Grenoble, France \\
$^{4}$ Department of Physics and Astronomy, University College London, Gower Str
eet, London, WC1E 6BT, UK \\
$^{5}$ IGN, Observatorio Astron\'omico Nacional, Calle Alfonso XIII, 28004 Madrid, Spain \\
$^{6}$ INAF, Istituto di Astrofisica e Planetologia Spaziali, via Fosso del Cavaliere 100, 00133 Roma, Italy \\
}

\begin{document}

\date{Accepted date. Received date; in original form date}

\pagerange{\pageref{firstpage}--\pageref{lastpage}} \pubyear{2011}

\maketitle

\label{firstpage}

\begin{abstract}
The formation of complex organic molecules
(COMs) in protostellar environments is a hotly debated topic. 
In particular, the relative importance 
of the gas phase processes as compared to a direct formation of COMs on 
the dust grain surfaces is so far unknown. 
We report here the first high-resolution images of acetaldehyde (CH$_3$CHO) emission towards
the chemically rich protostellar shock L1157-B1, obtained at  
2 mm with the IRAM Plateau de Bure interferometer.
Six blueshifted CH$_3$CHO lines with $E_{\rm u}$ = 26-35 K have been detected. 
The acetaldehyde spatial distribution follows  
the young ($\sim$ 2000 yr) outflow cavity produced by the impact of the jet with
the ambient medium, indicating that this COM is closely
associated with the region enriched by iced species evaporated from  
dust mantles and released into the gas phase. 
A high CH$_3$CHO relative abundance, 2--3 $\times$ 10$^{-8}$,
is inferred, similarly to what found in hot-corinos.   
Astrochemical modelling indicates that gas phase reactions  
can produce the observed quantity of acetaldehyde only if a large fraction
of carbon, of the order of 0.1\%, is locked into iced hydrocarbons.
\end{abstract}


\section{Introduction}

Complex organic molecules (COMs) have a key role among
the many molecules so far detected in space: since they follow the
same chemical rules of carbon-based chemistry, which terrestrial life
is based on, they may give us an insight into the universality of
life.  Of course, large biotic molecules are not detectable in space,
certainly not via (sub)millimeter observations.  However, to
determine whether pre-biotic molecules may form in
space, we first need to understand the basic mechanisms that form
smaller COMs.  There is an extensive literature on the subject and
still much debate on how COMs may form in space (e.g. Herbst \& van
Dishoeck 2009; Caselli \& Ceccarelli 2012; Bergin 2013).  Two basic
processes are, in principle, possible: COMs may form on the grain
surfaces or in gas phase.  It is possible and even probable that the
two processes are both important in different conditions for different
molecules.

Acetaldehyde
(CH$_3$CHO) has been detected in a large range of
interstellar conditions and with different abundances, namely in hot cores
(Blake et al. 1986), hot corinos (Cazaux et al. 2003), cold envelopes
(Jaber et al. 2014), Galactic Center clouds (Requena-Torres
et al. 2006) and pre-stellar cores (\"{O}berg et al. 2010).  Grain
surface models predict that CH$_3$CHO is one of the simplest COMs
and can be formed either by the combination of two radicals on the
grain surface, CH$_3$ and HCO, which become mobile when the grain
temperature reaches $\sim$30 K (Garrods \& Herbst 2006), or by
irradiation of iced CH$_4$, CO$_2$ and other iced species (Bennett et
al. 2005). For the former route, the two radicals are predicted to
be formed either because of the photolysis of more complex molecules
on the grain mantles or, more simply, because
of the partial hydrogenation of simple biatomic molecules on the grain
mantles (Taquet et al. 2012). Conversely, gas phase models claim that
acetaldehyde is easily formed by the oxidation of hydrocarbons, which
are produced by the hydrogenation of carbon chain on the grain mantles
(Charnley et al. 1992, 2004). Finally, a further possible
mechanism involving formation in the very high density gas-phase
immediately after ice mantles are sublimated has been proposed by
Rawlings et al. (2013).

In general, it is very difficult to distinguish which of these three
mechanisms are at work and, consequently, their relative
importance. 
The chemically rich shocked region L1157-B1 
offers a unique possibility to test these theories, 
as it is a place where the dust is not heated by
the protostar, but some of the grain mantles are 
sputtered/injected in the gas phase because of the passage of a shock  
(see e.g. Fontani et al. 2014).
The L1157-mm protostar ($d$ = 250 pc) drives a chemically rich outflow
(Bachiller et al. 2001), associated with molecular clumpy cavities
(Gueth et al. 1996), created by
episodic events in a precessing jet.  Located at the apex of the more
recent cavity, the bright bow shock called B1 has a kinematical age of
2000 years.  This shock spot has been the target
of several studies (e.g. the Large Programs
Herschel/CHESS\footnote{http://www-laog.obs.ujf-grenoble.fr/heberges/chess/}
(Chemical Herschel Surveys of Star forming regions; Ceccarelli et
al. 2010; and IRAM-30m/ASAI\footnote{http://www.oan.es/asai}
(Astrochemical Survey At IRAM).  In this Letter we report high spatial
resolution observations of acetaldehyde, with the aim to constrain and
quantify the contribution of gas phase chemistry to the CH$_3$CHO
formation.

\section{Observations}

L1157-B1 was observed with the IRAM Plateau de Bure (PdB) 
6-element array in April--May 2013 
using both the C and D configurations, 
with 21--176 m baselines, 
filtering out structures $\geq$ 20$\arcsec$, and providing
an angular resolution of 2$\farcs$5 $\times$ 2$\farcs$3 (PA =
90$\degr$).  The primary HPBW is $\sim 37\arcsec$.  The observed
CH$_3$CHO lines (see Table 1) at $\sim$ 134--136 GHz were detected
using the WideX backend which covers a 4 GHz spectral window at a 2 MHz
($\sim$ 4.4 km s$^{-1}$) spectral resolution.  The system temperature
was 100--200~K in all tracks, and the amount of precipitable water
vapor was generally $\sim$ 5~mm.  Calibration was carried out
following standard procedures, using
GILDAS-CLIC\footnote{http://www.iram.fr/IRAMFR/GILDAS}.  Calibration
was performed on 3C279 (bandpass), 1926+611, and 1928+738 (phase and
amplitude).  The absolute flux scale was set by observing MWC349
($\sim 1.5$~Jy at 134~GHz).  The typical rms noise in the 2 MHz
channels was $\sim$ 0.7 mJy beam$^{-1}$.

\section{Results: Images and Spectra}

Acetaldehyde emission has been clearly (S/N $\geq$ 10) detected
towards L1157--B1.  Fig. 1 shows the map of the
CH$_3$CHO(7$_{\rm 0,7}$--6$_{\rm 0,6}$) E and A lines integrated emission.
In order to verify whether the present CH$_3$CHO image is altered by
filtering of large-scale emission, we produced 
the CH$_3$CHO(7$_{\rm 0,7}$--6$_{\rm 0,6}$) E+A spectrum
summing the emission measured at PdBI in a circle of
diameter equal to the half-power beam width (HPBW) of the IRAM-30m
telescope (17$\arcsec$). We evaluated the missing flux by
comparing such emission with the spectrum directly measured with the single-dish
(from the ASAI spectral survey, Lefloch et al., in
preparation). 
As already found for HDCO by Fontani et al. (2014),
with the PdBI we recover more than 80\% of the flux,
indicating that both
tracers do not have significant extended structures.
The spatial distribution reported in Fig. 1 
shows that CH$_3$CHO is mainly
associated with two regions: (i) the eastern B0-B1 cavity opened
by the precessing jet (called `E-wall', see Fig. 1 in Fontani et
al. 2014), and (ii) the arch-like structure composed
by the B1a-e-f-b clumps identified by CH$_3$CN (called `arch'). The
red, turquoise, and magenta polygons shown in Fig. 1 sketch out these two regions,
intersecting at the position of the B1a clump. Note that B1a is in
turn located where the precessing jet is expected to impact the cavity
producing a dissociative J-shock (traced by high velocity SiO, H$_2$O,
[FeII], [OI], and high-J CO emission: e.g. Gueth et al. 1998,
Benedettini et al. 2012). 

\begin{figure}
\begin{minipage}{62mm}
\resizebox{\hsize}{!}{\includegraphics[angle=0]{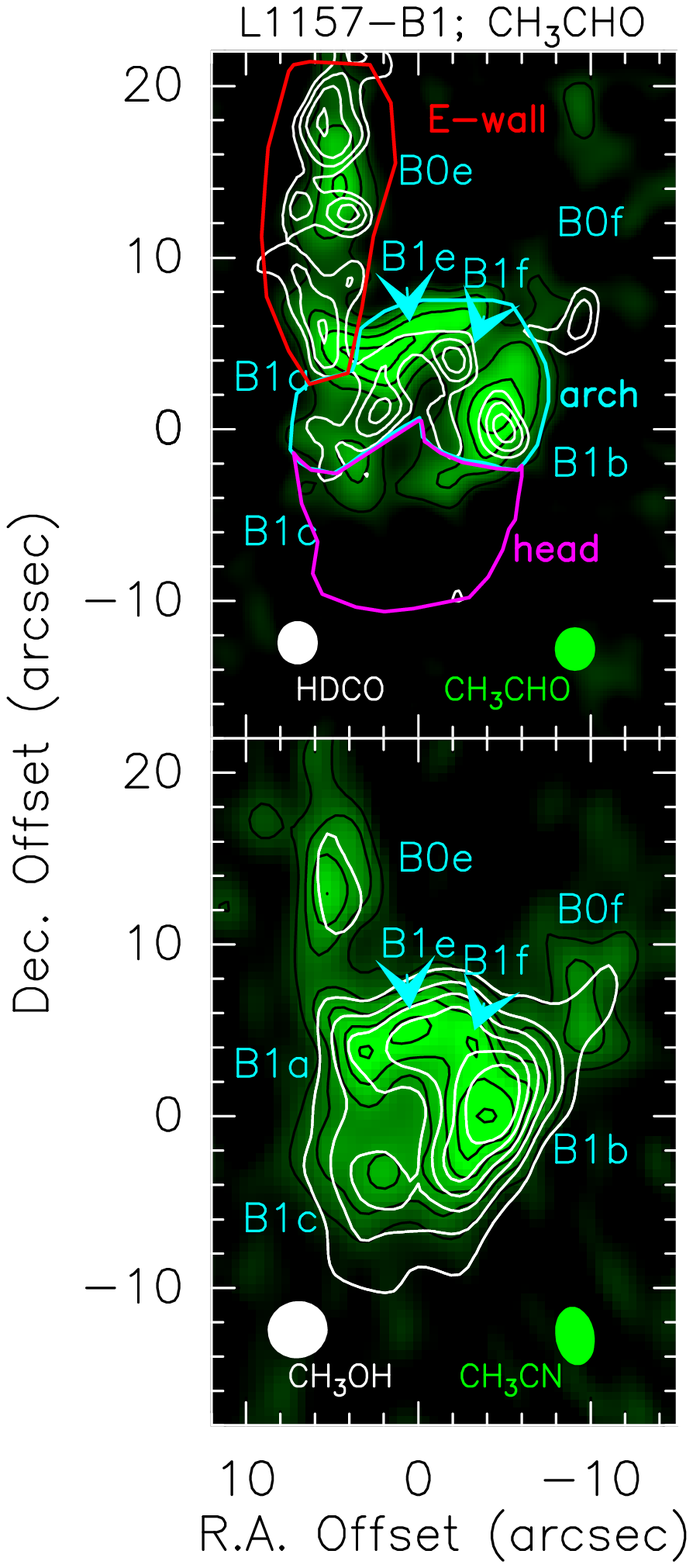}}
 \end{minipage}
 \caption[]
{Chemical differentation in L1157--B1: the maps are centred at:
$\alpha({\rm J2000})$ = 20$^h$ 39$^m$ 09$\fs$5,
$\delta({\rm J2000})$ = +68$\degr$ 01$\arcmin$ 10$\farcs$0, i.e. at
$\Delta\alpha$ = +21$\fs$7 and $\Delta\delta$ = --64$\farcs$0 from the driving protostar.
{\it Upper panel:} CH$_3$CHO(7$_{0,7}$--6$_{0,6}$)E+A integrated emission 
(green colour, black contours) on top
of the HDCO(2$_{1,1}-1_{0,1}$) line (white contours; Fontani et al. 2014).
First contour and steps of the CH$_3$CHO image correspondes to 3$\sigma$ (1 mJy beam $^{-1}$).
The ellipses show the 
synthesised HPBW ($2\farcs5\times2\farcs3$, PA = 90$\degr$).
The red, turquoise, and magenta polygons called 'E-wall', 'arch', and 'head' indicate the
3 portions of L1157-B1 selected by Fontani et al. (2014) to
investigate H$_2$CO deuteration.
{\it Bottom panel:} CH$_{3}$CN(8$_{K}$--7$_{K}$)
emission (green colour, black contours; Codella et al. 2009)
on top of the CH$_3$OH(3$_{2}$--2$_{K}$) emission (white; Benedettini et al. 2013).
The HPBWs are: $3\farcs4\times2\farcs1$ (PA = 10$\degr$) for CH$_{3}$CN and
$3\farcs5\times2\farcs3$ (PA = 12$\degr$) for CH$_3$OH.
The labels indicate the L1157-B1 clumps identified using the CH$_{3}$CN image (Codella et al. 2009).}
\end{figure}

Figure 2 shows the CH$_3$CHO line spectrum observed with the 3.6 GHz
WideX bandwidth towards the brightest clump, B1a. Up to six lines
($E_{\rm u}$ = 26--35 K, see Table 1) are detected
with a S/N $>$ 3. Using the GILDAS--Weeds package (Maret et al. 2011)
and assuming optically thin emission and LTE conditions, we produced
the synthetic spectrum (red line in Fig. 2) that best fits the
observed one. Note that the CH$_3$CHO lines are blue-shifted,
by 2 km s$^{-1}$, with respect to the cloud systemic velocity (+2.6
km s$^{-1}$: Bachiller \& Per\'ez Guti\'errez 1997), and have
linewidths of 8 km s$^{-1}$. 
Similarly, we extracted the CH$_3$CHO
line spectrum towards the three B1 zones, `E-wall', `arch', and
`head', shown in Fig. 1. Table 2 reports the measured peak
velocities, intensities (in $T_{\rm B}$ scale), FWHM linewidths, and
integrated intensities, for each of the three zones. 

\begin{table}
  \caption{List of CH$_3$CHO transitions detected towards L1157-B1}
  \begin{tabular}{lcccc}
  \hline
\multicolumn{1}{c}{Transition} &
\multicolumn{1}{c}{$\nu$$^{\rm a}$} &
\multicolumn{1}{c}{$E_{\rm u}$$^a$} &
\multicolumn{1}{c}{$S\mu^2$$^a$} &
\multicolumn{1}{c}{log(A/s$^{-1}$)$^a$} \\
\multicolumn{1}{c}{ } &
\multicolumn{1}{c}{(GHz)} &
\multicolumn{1}{c}{(K)} &
\multicolumn{1}{c}{(D$^2$)} & 
\multicolumn{1}{c}{} \\
\hline
(7$_{\rm 0,7}$--6$_{\rm 0,6}$)E  & 133.830 & 26 & 88.5 & --4.04\\
(7$_{\rm 0,7}$--6$_{\rm 0,6}$)A  & 133.854 & 26 & 88.4 & --4.08 \\
(7$_{\rm 2,6}$--6$_{\rm 2,5}$)A  & 134.694 & 35 & 81.3 & --4.11 \\
(7$_{\rm 2,6}$--6$_{\rm 2,5}$)E  & 134.895 & 35 & 79.7 & --4.12 \\
(7$_{\rm 2,5}$--6$_{\rm 2,4}$)E  & 135.477 & 35 & 79.7 & --4.11 \\
(7$_{\rm 2,5}$--6$_{\rm 2,4}$)A  & 135.685 & 35 & 81.3 & --4.10 \\
\hline
\end{tabular}

$^a$ From the Jet
Propulsion Laboratory database (Pickett et al. 1998). \\
\end{table}

\begin{table*}
  \caption{Observed parameters (in $T_{\rm B}$ scale) of the 
CH$_3$CHO(7$_{\rm 0,7}$--6$_{\rm 0,6}$)E and A emission,
and acetaldehyde column densities $N_{\rm CH_3CHO}$ derived
in the 3 regions identified in Fig.~1 (E-wall, arch, and head) following
Fontani et al. (2014), see Sect. 3.
The (range of) excitation temperatures ($T_{\rm ex}$) used
to derive $N_{\rm CH_3CHO}$ have been assumed equal to the rotation temperatures
derived in Codella et al. (2012), Lefloch et al. (2012), and Fontani et al. (2014).
The last columns report the $X(CH_3CHO)$/$X(CH_3OH)$ 
and $X(CH_3CHO)$/$X(HDCO)$ abundance ratios
using the CH$_3$OH and HDCO data (and similar beams) by Benedettini et al. (2013) 
and Fontani et al. (2014).}
  \begin{tabular}{lccccccc}
  \hline
\multicolumn{1}{c}{Transition} &
\multicolumn{1}{c}{$T_{\rm peak}$$^a$} &
\multicolumn{1}{c}{$V_{\rm peak}$$^a$} &
\multicolumn{1}{c}{$FWHM$$^a$} &
\multicolumn{1}{c}{$F_{\rm int}$$^a$} &
\multicolumn{1}{c}{$N_{\rm CH_3CHO}$$^b$} &
\multicolumn{1}{c}{CH$_3$CHO/CH$_3$OH} &
\multicolumn{1}{c}{CH$_3$CHO/HDCO} \\ 
\multicolumn{1}{c}{} &
\multicolumn{1}{c}{(mK)} &
\multicolumn{1}{c}{(km s$^{-1}$)} &
\multicolumn{1}{c}{(km s$^{-1}$)} &
\multicolumn{1}{c}{(mK km s$^{-1}$)} &
\multicolumn{1}{c}{(10$^{12}$ cm$^{-2}$)} &
\multicolumn{1}{c}{(10$^{-2}$)} &
\multicolumn{1}{c}{ } \\ 
\multicolumn{1}{c}{ } &
\multicolumn{1}{c}{ } &
\multicolumn{1}{c}{ } &
\multicolumn{1}{c}{ } &
\multicolumn{1}{c}{ } &
\multicolumn{1}{c}{10 K -- 70 K} &
\multicolumn{1}{c}{10 K -- 70 K} &
\multicolumn{1}{c}{10 K -- 70 K } \\
\hline
\multicolumn{8}{c}{E-wall} \\
%
7$_{\rm 0,7}$--6$_{\rm 0,6}$ E & 30(3) & +0.4(0.4) & 8.2(1.0) & 264(28) & & & \\ [-1ex]
 & &  &  & & 5.0(0.3)--9.2(0.5) & 1.7(0.1)--11.0(0.6) & 1.9(0.2)--0.9(0.1) \\ [-1.1ex]
7$_{\rm 0,7}$--6$_{\rm 0,6}$ A & 29(3) & +0.5(0.5) & 8.6(1.3) & 263(31) &  &  &  \\
\multicolumn{8}{c}{Arch} \\
7$_{\rm 0,7}$--6$_{\rm 0,6}$ E & 71(7) & --0.9(0.3) & 9.2(0.8) & 748(54) & & & \\ [-1ex] 
 & &  &  & & 15.9(0.1)--29.4(0.1) & 0.6(0.1)--4.2(0.1) & 7.6(1.1)--3.7(0.5) \\ [-1.1ex]
7$_{\rm 0,7}$--6$_{\rm 0,6}$ A & 76(7) & --0.2(0.9) & 8.0(0.5) & 608(42) & & &  \\  
\multicolumn{8}{c}{Head} \\
7$_{\rm 0,7}$--6$_{\rm 0,6}$ E  & 14(3) & +0.9(0.5) & 8.0(1.1) & 81(13) & & & \\ [-1ex] 
 & &  &  & & 1.7(0.3)--3.1(0.5) & 0.2(0.1)--0.7(0.1) &  $\geq$ 1.5 \\ [-1.1ex]
7$_{\rm 0,7}$--6$_{\rm 0,6}$ A & 13(3) & +0.9(0.6) & 5.2(1.6) & 81(13) &  &  &  \\  
\hline
\end{tabular}

$^a$ The errors are the gaussian fit uncertainties. The spectral resolution is 4.4 km s$^{-1}$.
$^b$ Derived using the (7$_{\rm  0,7}$--6$_{\rm 0,6}$) E and A emissions.
\end{table*}

\begin{figure*}
\begin{minipage}{150mm}
 \resizebox{\hsize}{!}{\includegraphics[angle=0]{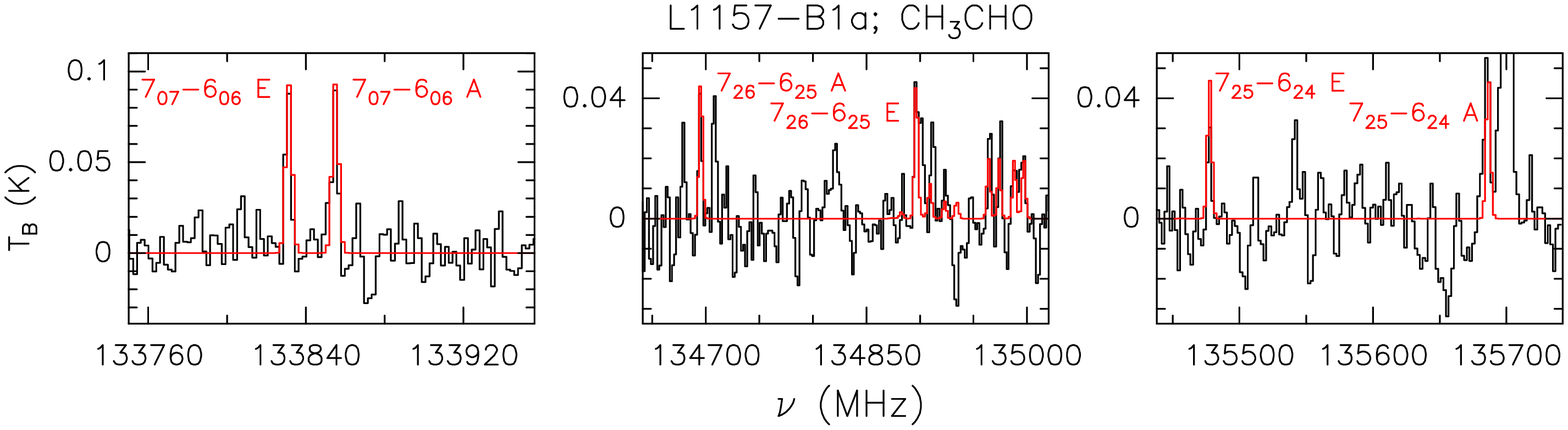}}
\end{minipage}
 \caption[]
{CH$_3$CHO emission (in $T_{\rm B}$ scale) extracted at the B1a position
($\alpha({\rm J2000})$ = 20$^h$ 39$^m$ 10$\fs$2, $\delta({\rm J2000})$ = +68$\degr$ 01$\arcmin$ 12$\farcs$0).
the three panels show the frequency intervals of the 4 GHz wide WideX where the six CH$_3$CHO lines
with S/N $\geq$ 3$\sigma$ (33 mK) are located (see Table 1). 
The red line shows the synthetic spectra which better reproduce the observations: 
it has been obtained with the GILDAS--Weeds package (Maret et al. 2011) 
assuming optically thin emission and LTE conditions with
$N_{\rm CH_3CHO}$ = 9 $\times$ 10$^{13}$ cm$^{-2}$, $T_{\rm ex}$ = 15 K, $v_{\rm LSR}$ = +0.6 km s$^{-1}$,
and FWHM linewidth = 8.0 km s$^{-1}$.} 
\end{figure*}

\section{CH$_3$CHO abundance}

Figure 1 compares
the CH$_3$CHO distribution with that of HDCO (Fontani et
al. 2014), showing an excellent agreement, with weak
or no emission at the head of the bow B1 structure (called `head').
The acetaldehyde emission is
concentrated towards the `E-wall' and `arch' zones, namely the
part of B1 associated with the most recent shocked material, as probed
by the HDCO emission. This is further
supported by the fact that the brightest acetaldehyde emission comes
where also CH$_3$OH, another dust mantle product, and CH$_3$CN, a 6-atoms
COM, emission peak (Codella et al. 2009, Benedettini et
al. 2013). Finally, the CH$_3$CHO observed emission is also confined
in the low-velocity range ($FWHM$ $\sim$ 8 km s$^{-1}$) of the L1157-B1 outflow, which is dominated
by the extended B1 bow-cavity, according to Lefloch et al. (2012) and
Busquet et al. (2014). In summary, similarly to HDCO, 
CH$_3$CHO traces the extended interface between the
shock and the ambient gas, which is chemically enriched by the
sputtering of the dust mantles.

To derive the column density, we used the LTE populated and
optically thin assumption and best fitted the six detected lines of
Tables 1--2. Towards the B1a peak, we find
$N_{\rm CH_3CHO}$ = 9 $\times$ 10$^{13}$ cm$^{-2}$,
and a rotational temperature of $T_{\rm rot}$ = 15 K, in agreement
with the value derived for the molecular cavity from single-dish CO
and HDCO measurements (10--70 K; Lefloch et al. 2012,
Codella et al. 2012). Assuming
rotational temperatures between 10 and 70 K (Table 2) we derived 
a column density of 5--30 $\times$ 10$^{12}$ cm$^{-2}$ in the `E-wall' and `arch' regions, 
and $\sim$ 2--3 10$^{12}$ cm$^{-2}$ in the `head'.
The size of the regions (at 3$\sigma$ level) is 9$\arcsec$ (`E-wall'), 
7$\arcsec$ (`arch'), and 8$\arcsec$ (`head').
An estimate of the CH$_3$CHO abundance can be derived using the
the CO column density $\simeq$ 10$^{17}$ cm$^{-2}$ derived by Lefloch et al. (2012)  
on a 20$\arcsec$ scale. 
We derived $N_{\rm CH_3CHO}$ using the CH$_3$CHO spectrum 
extracted on the same scale and assuming 10--70 K.
We find $N_{\rm CH_3CHO}$ $\sim$ 0.9--1.6 $\times$ 10$^{13}$
cm$^{-2}$, which implies a high abundance, $X({\rm CH_3CHO})$
$\simeq$ 2--3 $\times$ 10$^{-8}$, similar to what has been measured in
hot-corinos ($\simeq$ 2--6 $\times$ 10$^{-8}$, Cazaux et al. 2003),
and larger than that measured in prestellar cores ($\sim$ 10$^{-11}$,
Vastel et al. 2014) and towards high-mass star forming regions ($\sim$
10$^{-11}$--10$^{-9}$, Cazaux et al. 2003; Charnley 2004).

\section{Gas phase formation of CH$_3$CHO}

The ratio between $N_{\rm CH_3CHO}$ 
and the column density of HDCO, i.e. a molecule which 
in L1157-B1 is predominantly released by grain mantles (Fontani et al. 2014), 
is higher (even if we consider the uncertainties, see Table 2) in  
the `arch' with respect to the `E-wall' by 
a factor $\sim$ 2--8. Assuming the same grain mantle composition 
and release mechanism, this difference suggests that, in the `arch',
a significative fraction of the observed CH$_3$CHO is formed in 
the gas phase.
In the gas phase, the injection from grain mantles of ethane (C$_2$H$_6$)  
is expected to drive first C$_2$H$_5$ and 
successively acetaldehyde  
(e.g. Charnley 2004; Vasyunin \& Herbst 2013):
the overlap between the HDCO (Fontani et al.
2014) and CH$_3$CHO emitting regions supports this scenario.
We can, therefore, use the measured CH$_3$CHO abundance to constrain
the quantity of C$_2$H$_5$ that has to be present in the gas phase 
in order to produce
the observed quantity of CH$_3$CHO.
To this end, we use the chemical code 
ASTROCHEM\footnote{http://smaret.github.com/astrochem/},
a pseudo time
dependent model that follows the evolution of a gas cloud with a fixed
temperature and density considering a network of chemical reactions in the gas phase.
We followed the
same 2-steps procedure adopted in Podio et al. (2014) and Mendoza et al. (2014),
to first compute the steady-state abundances in the cloud 
(i.e. T$_{\rm kin}$=10 K, n$_{\rm H_2}$=10$^4$ cm$^{-3}$, $\zeta$=3 10$^{16}$ 
s$^{-1}$); and then we follow the gas evolution over 2000 yr at the shocked
conditions (i.e. T$_{\rm kin}$=70 K and n$_{\rm H_2}$=10$^5$
cm$^{-3}$). To estimate the influence of a possibly larger
gas T$_{\rm kin}$ during the passage of the shock, we also
run cases with temperatures up to 1000 K. 
We adopt the
OSU\footnote{http://faculty.virginia.edu/ericherb}
chemical network and assume visual extinction of A$_{\rm V}$ = 10 mag
and grain size of 0.1 $\mu$m.  
We assume that the abundances of OCS and CO$_2$ are also
enhanced by the passage of the shock, namely their abundance in step 2
is $X({\rm CO_2})$ = 6 10$^{-5}$ and $X({\rm OCS})$ = 6 10$^{-6}$.
Similarly, we assume that the abundance of methanol in step 2 is 2
10$^{-6}$, in agreement with the most recent determination in L1157-B1
by Mendoza et al. (2014). Finally, we varied the C$_2$H$_5$ abundance
from 2 $\times$ 10$^{-7}$ to 2 $\times$ 10$^{-5}$.
As expected, the predicted steady-state abundance of acetaldehyde in
the cloud is very low (1.5 10$^{-15}$). However, once C$_2$H$_5$ is
in the gas phase, it rapidly reacts with
oxygen forming abundant acetaldehyde on timescale shorter than 100
years (Fig. 3). The CH$_3$CHO abundance reaches the observed value,
$\simeq$ 2--3 $\times$ 10$^{-8}$, at the shock age (2000 years), for
C$_2$H$_5$ $\sim$ 2--6 $\times$ 10$^{-7}$. Note that we obtain the
same result if the gas temperature is $\leq 500$ K, and a
30\% higher value at 1000 K.
Figure 3 shows also that 
the CH$_3$CHO/CH$_3$OH abundance ratio is expected to
drop between 10$^3$ yr and 10$^4$ yr. A different age could,
therefore, justify the slightly smaller CH$_3$CHO/CH$_3$OH ratio
observed towards the `head' region.

\begin{figure}
\begin{minipage}{85mm}
 \resizebox{\hsize}{!}{\includegraphics[angle=0]{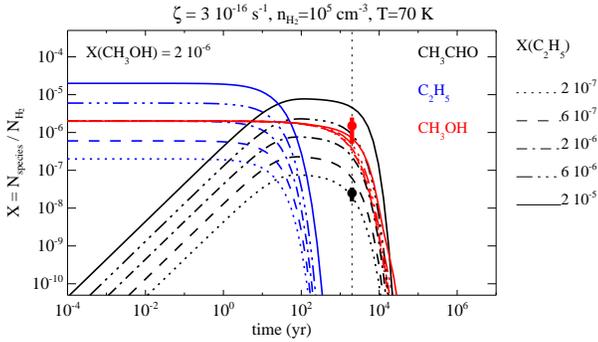}}
\end{minipage}
 \caption[]
{Evolution of acetaldehyde (CH$_3$CHO, black), (C$_2$H$_5$, blue), and methanol (CH$_3$OH, red) 
abundances in the shock as a function of time. Observed abundances (colour circles) are 
overplotted at the shock age ($t_{shock} \sim 2000$ years, vertical dotted line). The evolution 
is computed from steady-state values in the cloud ($n_{H_2} = 10^4$ cm$^{-3}$, $T_{kin} = 10$ K, 
$\zeta = 3 \times 10^{-16}$ s$^{-1}$) by enhancing the gas temperature and density 
($n_{H_2} = 10^5$ cm$^{-3}$, $T_{kin} = 70$ K), and the abundance of molecules which are thought to 
be sputtered off dust grain mantles. We set $X_{CO_2} = 6 \times 10^{-5}$ and
$X_{OCS} = 6 \times 10^{-6}$ as in Podio et al. 2014, $X_{CH_3OH} = 2 \times 10^{-6}$  
(Mendoza et al. 2014), and vary the abundance of C$_2$H$_5$ between $2 \times 10^{-7}$ and 
$2 \times 10^{-5}$.}
\end{figure}

\section{Discussion and conclusions}

We have shown that acetaldehyde is abundant, $X({\rm CH_3CHO})$
$\simeq$ 2--3 $\times$ 10$^{-8}$, in the gas associated with the
passage of a shock and enriched by iced species sputtered from grain
mantles and injected into the gas phase. The measured acetaldehyde
abundance could be consistent with the scenario of oxydation of gaseous
hydrocarbons formed in a previous phase and released by the grain
mantles. However, the abundance of the C$_2$H$_5$ required to
reproduce the measured CH$_3$CHO is very high, $\sim$ 2--6 $\times$ 10$^{-7}$,
namely less than 0.6\% the elemental gaseous carbon.
There are no observations of C$_2$H$_5$, hence
it is impossible to compare with direct estimates of the
abundance of this molecule. However, it has been argued
that large quantities of frozen methane, of a few \% of iced mantle
water, is found around the L1527-mm protostar, where the detection
of CH$_3$D (Sakai et al. 2012) indicates  
$X({\rm CH_4})$ $\simeq$ 0.4--1.5 $\times$ $10^{-5}$.
This large abundance has
been attributed to a low density of the pre-collapse core from which
L1527-mm originated (Aikawa et al. 2008). Interestingly, the analysis of the
deuteration of water, methanol and formaldehyde in L1157-B1 led
Codella et al. (2012) to conclude that also the mantles of L1157-B1
were formed in relatively low density ($\sim$ 10$^3$ cm$^{-3}$)
conditions.

To conclude, in the specific case of L1157-B1, gas phase reactions 
can produce the observed quantity of acetaldehyde only if a large fraction 
of carbon, of the order of 0.1\%, is locked into iced hydrocarbons. 
Further observations of the hydrocarbons abundance in 
L1557-B1 are needed to confirm or dismiss our hypothesis.

\section*{Acknowledgments}

The authors are grateful to P. Caselli for instructive comments and
suggestions, as well as to the IRAM staff for its help in the
calibration of the PdBI data. 
This research 
has received funding from the European Commission Seventh Framework
Programme (FP/2007-2013, n. 283393, RadioNet3), 
the PRIN INAF 2012 -- JEDI, and 
the Italian Ministero dell'Istruzione, Universit\`a e Ricerca through
the grant Progetti Premiali 2012 -- iALMA. 
LP has received funding from the European Union Seventh Framework 
Programme (FP7/2007-2013, n. 267251).
CC and BL acknowledge the
financial support from the French Space Agency CNES, and RB
from Spanish MINECO (FIS2012-32096).

{}

\bsp

\label{lastpage}

\end{document}